\documentclass[reprint,superscriptaddress,amsmath,amssymb,aps,pra,longbibliography]{revtex4-2}
\usepackage{graphicx}% Include figure files
\usepackage{dcolumn}% Align table columns on decimal point
\usepackage{bm}% bold math
\usepackage{braket}
\usepackage{xcolor}
\usepackage[
    colorlinks=true,
    linkcolor=blue,
    citecolor=blue,
    urlcolor=blue,
    bookmarks=true,
    breaklinks=true
]{hyperref}% add hypertext capabilities
\usepackage[all]{hypcap}
\definecolor{ColorName}{RGB}{18,220,168}

\begin{document}
\preprint{APS/123-QED}

%\title{Quantum Criticality and Ground State Factorization\\ in Competing Coherent Rydberg Chain}% Force line breaks with \\
% \thanks{A footnote to the article title}%

\title{Quantum criticality and factorization in a constrained Rydberg spin chain}
\author{Yuan Jiang}
\affiliation{School of Systems Science, Beijing Normal University, Beijing 100875, China}

\author{Wen-Long You}
\email{wlyou@nuaa.edu.cn}
\affiliation{Center for the Cross-disciplinary Research of Space Science and Quantum-technologies (CROSS-Q), College of Physics, Nanjing University of Aeronautics and Astronautics, Nanjing 211106, China}
%\affiliation{Key Laboratory of Aerospace Information Materials and Physics (NUAA), MIIT, Nanjing 211106, China}

\author{Liangsheng Li}
\email{liliangshengbititp@163.com}
\affiliation{National Key Laboratory of Scattering and Radiation, Beijing 100854, China}

\author{Maoxin Liu}
\email{mxliu@bnu.edu.cn}
\affiliation{School of Systems Science, Beijing Normal University, Beijing 100875, China}

\date{\today}

\begin{abstract}
    We investigate the zero-temperature phase diagram of a one-dimensional constrained quantum spin chain realized in coherently driven Rydberg atom arrays with competing local Rabi driving and dipole-dipole exchange interactions.
    Projecting onto the blockade-constrained Hilbert space yields an effective model in which local and nonlocal quantum fluctuations compete on equal footing.
    Combining exact diagonalization, density matrix renormalization group, and variational uniform matrix product state calculations, we establish a complete phase diagram comprising a Luttinger liquid, an antiferromagnetic ordered phase, and a polarized paramagnet.
    We identify two distinct mechanisms for the destruction of antiferromagnetic order: a conventional Ising transition at strong driving and a continuous quantum melting into the Luttinger liquid at weak driving, characterized using entanglement-based diagnostics and finite-entanglement scaling.
    In addition, we uncover an exact ground-state factorization line embedded within the ordered phase, providing an analytically tractable zero-entanglement reference point for experiments with programmable Rydberg quantum simulators.
\end{abstract}

%\keywords{Suggested keywords}%Use showkeys class option if keyword
%display desired
\maketitle

%\tableofcontents
\section{\label{sec:introduction}INTRODUCTION}

In recent years, experimental advances in quantum simulation with arrays of individually trapped neutral atoms have enabled access to strongly correlated many-body physics~\cite{weimerRydbergQuantumSimulator2010,mukherjeeManybodyPhysicsAlkalineearth2011,lowExperimentalTheoreticalGuide2012,bernienprobingManybodyDynamics2017,henrietQuantumComputingNeutral2020,morgadoQuantumSimulationComputing2021}. Among these platforms, Rydberg atom arrays~\cite{bakrQuantumGasMicroscope2009,saffmanQuantumInformationRydberg2010, shersonSingleatomresolvedFluorescenceImaging2010,labuhnTunableTwodimensionalArrays2016,ecknerRealizingSpinSqueezing2023} —typically implemented with reconfigurable optical tweezers~\cite{endresAtombyatomAssemblyDefectfree2016} — are particularly distinguished by their strong interactions and the Rydberg blockade mechanism. In the strong-blockade regime, the blockade imposes nontrivial local exclusion rules that project the system dynamics onto a constrained Hilbert space, thereby realizing constrained spin and lattice-gas models with emergent phenomena including ordered phases~\cite{ebadiQuantumPhasesMatter2021,schollQuantumSimulation2D2021}, emergent gauge structures~\cite{celiEmergingTwoDimensionalGauge2020,chengEmergentGaugeTheory2024}, and nonergodic scarred dynamics~\cite{turnerWeakErgodicityBreaking2018,turnerQuantumScarredEigenstates2018,schecterWeakErgodicityBreaking2019,jamesNonthermalStatesArising2019,serbynQuantumManybodyScars2021,moudgalyaQuantumManybodyScars2022,iadecolaQuantumManybodyScar2020,suObservationManybodyScarring2023}.

Most existing studies, however, have focused on single-Rydberg-state implementations, where interactions are predominantly short-ranged and the effective models are limited in structure. Recent experimental work has extended this paradigm by coherently coupling two Rydberg levels~\cite{maxwellStorageControlOptical2013,teixeiraMicrowavesProbeDipole2015,baurSinglePhotonSwitchBased2014,bettelliExcitonDynamicsEmergent2013,tiarksSinglephotonTransistorUsing2014,gorniaczykSinglePhotonTransistorMediated2014,faheyImagingDipoledipoleEnergy2015}, thereby realizing the simultaneous presence of short-range van der Waals (vdW) interactions and long-range resonant dipolar exchange~\cite{huberDipoleInteractionMediatedLaserCooling2012,zhaoAtomicRydbergReservoirs2012,barredoCoherentExcitationTransfer2015,browaeysExperimentalInvestigationsDipole2016,abumwisExtendedCoherentlyDelocalized2020,browaeysManybodyPhysicsIndividually2020,zeybekQuantumPhasesCompeting2023}. This setup leads to an effective constrained spin model featuring both transverse fields and power-law spin-exchange interactions, which cannot be realized using single-Rydberg-state. Experiments have demonstrated coherent excitation transport~\cite{gunterObservingDynamicsDipoleMediated2013}, spin squeezing~\cite{bornetScalableSpinSqueezing2023}, continuous symmetry breaking~\cite{chenContinuousSymmetryBreaking2023}, supersolidity~\cite{homeierSupersolidityRydbergTweezer2025}, and engineered topological transport~\cite{sunGroundStateManipulation2025}. The coexistence of strong local constraints and long-range interactions makes this platform particularly well-suited for exploring quantum criticality and competing orders in low dimensions, necessitating a systematic theoretical understanding.

Despite these advances, theoretical descriptions remain incomplete. On the one hand, studies focus on the exactly resonant limit, where dipolar exchange is used to engineer XXZ models~\cite{schollMicrowaveEngineeringProgrammable2022} or to explore the order-by-disorder mechanism~\cite{zeybekFormationRydbergCrystals2025}. On the other hand, research including finite detuning often lacks Rabi driving, focusing on integrability and Hilbert space fragmentation~\cite{bastianelloFragmentationEmergentIntegrable2022,PhysRevResearch.7.023099}. In experimentally relevant regimes, however, finite detuning is routinely employed to control excitation density while maintaining coherent Rabi driving, a setup which has already been realized in experiments~\cite{deleseleucobservationsymmetryprotectedtopological2019}. This places the system deep in the strong-blockade regime, where a low-energy effective description in terms of constrained quantum spin models is more appropriate. However, a systematic characterization of the resulting phase diagram and critical behavior, where longitudinal detuning and transverse driving compete on equal footing, has been missing.

Here, we fill this gap by determining the complete zero-temperature phase diagram of a one-dimensional constrained quantum spin model generated by two coherently coupled Rydberg states at arbitrary detuning and driving strength. Using a combination of numerical and analytical approaches, we find that weak driving stabilizes a Luttinger liquid (LL) described by an effective Tomonaga–Luttinger theory, intermediate driving gives rise to an Ising antiferromagnetic (AFM) phase stabilized by the interplay of local transverse fields, blockade-induced constraints, and long-range dipolar exchange, while strong driving favors a fully polarized paramagnetic (PM) phase. The transition between the AFM and PM phases is associated with spontaneous breaking of discrete translational symmetry, whereas the melting of crystalline order into the LL proceeds via a continuous quantum phase transition consistent with LL criticality. Finally, we identify an exact ground-state factorization arising from destructive interference between the projected Rabi drive and dipolar exchange processes, which yields an analytically tractable point that constrains the surrounding strongly correlated regime.

\section{\label{sec:model}MODEL AND METHODS}

\begin{figure}[t]
    \centering
    \includegraphics[width=\columnwidth]{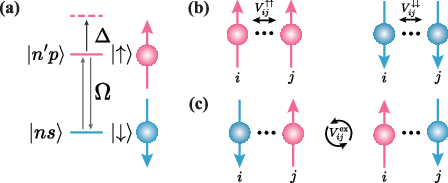}
    \caption{\label{fig:Fig1}
        (a) Energy level scheme of a single atom. A microwave field with Rabi frequency $\Omega$ and detuning $\Delta$ couples two Rydberg states, mapped to a pseudo-spin-1/2 system $\{|\!\downarrow\rangle, |\!\uparrow\rangle\}$.
        (b) Diagonal vdW interaction strengths $V_{ij}^{\uparrow\uparrow}$ and $V_{ij}^{\downarrow\downarrow}$ between atoms at sites $i$ and $j$.
        (c) Off-diagonal resonant dipole-dipole coupling strength $V_{ij}^{\rm ex}$ driving the spin exchange process.}
    \makeatletter
    \def\@currentlabel{\thefigure(a)}\label{fig:Schematic_a}
    \def\@currentlabel{\thefigure(b)}\label{fig:Schematic_b}
    \def\@currentlabel{\thefigure(c)}\label{fig:Schematic_c}
    \makeatother
\end{figure}

We consider a one-dimensional array of trapped neutral atoms, where each site $i$ is modeled as a two-level system. This system consists of a lower-lying Rydberg state $|ns\rangle_i$ and a higher-lying Rydberg state $|n'p\rangle_i$,  with $n$ and $n'$ denoting the principal quantum numbers. As illustrated in Fig.~\ref{fig:Schematic_a}, a global microwave field with Rabi frequency $\Omega$ and detuning $\Delta$ couples these two levels, effectively forming a pseudospin-1/2 degree of freedom where $|ns\rangle_i \text{ maps to } \ket{\downarrow}_i$ and $|n'p\rangle_i \text{ maps to } \ket{\uparrow}_i$. This specific choice of Rydberg state pair endows the system with two distinct types of atomic interactions that differ in their physical character, distinguished by spatial decay with respect to the inter-atomic distance $a|i-j|$, and $a$ is the lattice spacing: (i) the diagonal vdW interactions between atoms in the same Rydberg levels, which scale as $1/(a|i-j|)^6$ with interaction strengths (see Fig.~\ref{fig:Schematic_b}) $V^{\downarrow\downarrow}_{ij} = C^{\downarrow\downarrow}_6/(a|i-j|)^6$ and $V^{\uparrow\uparrow}_{ij} = C^{\uparrow\uparrow}_6/(a|i-j|)^6$, where $C^{\sigma\sigma}_6$ are the corresponding dispersion coefficients, and (ii) the off-diagonal dipole-dipole interaction, which scales as $1/(a|i-j|)^3$ with strength (see Fig.~\ref{fig:Schematic_c}) $V^{\rm ex}_{ij} = C_3/(a|i-j|)^3$, where $C_3$ is the dipolar exchange coefficient, drives a coherent state exchange $|ns\rangle_i |n'p\rangle_j \leftrightarrow |n'p\rangle_i |ns\rangle_j$. Considering all these contributions, the full Hamiltonian is given by:
\begin{eqnarray}
%\begin{split}
\hat{H}_{\rm full} &= & \sum\nolimits_i (\Omega \hat{\sigma}_i^x + \Delta \hat{n}_i )+ \! \sum\nolimits_{i<j} \! V^{\rm ex}_{ij}(\hat{\sigma}_i^+ \hat{\sigma}_{j}^- \! + \text{H.c.}) \nonumber \\
& +& \! \sum\nolimits_{i<j} \! [ V^{\uparrow\uparrow}_{ij} \hat{n}_i \hat{n}_j \! + \! V^{\downarrow\downarrow}_{ij} (1-\hat{n}_i) (1-\hat{n}_j) ],
%\end{split}
\label{eq:Hfull}
\end{eqnarray}
where the operators are defined in terms of the elementary transitions $\hat{\sigma}_i^{\alpha\beta} = |\alpha\rangle_i \langle \beta|_i$ between the Rydberg basis states $\alpha, \beta \in \{|ns\rangle, |n'p\rangle\}$. Specifically, we identify $\hat{n}_i \equiv \hat{\sigma}_i^{n'p, n'p}$, $\hat{\sigma}_i^+ \equiv \hat{\sigma}_i^{n'p, ns} = (\hat{\sigma}_i^-)^\dagger$, and $\hat{\sigma}_i^x \equiv \hat{\sigma}_i^+ + \hat{\sigma}_i^-$.

To derive an effective description, we first absorb the static long-range $\ket{\downarrow\downarrow}$ background to a site-dependent effective detuning $\Delta'_i = \Delta - \sum_{j \neq i} V^{\downarrow\downarrow}_{ij}$. This expression is obtained by expanding the $(1-n_i)(1-n_j)$ terms and collecting the linear contributions from the double summation $\sum_{i<j}$, which yields a single summation over $j \neq i$ for each site $i$. In practice, the detuning $\Delta$ is tuned to compensate for this background offset, ensuring that $\Delta'_i$ remains finite and controllable. Exploiting the rapid spatial decay of interactions, we truncate all terms beyond nearest neighbors (NN), yielding the NN Hamiltonian:
\begin{eqnarray}
        \hat{H}_{\rm NN} &=& \! \sum_i   (\Omega \hat{\sigma}_i^x \! + \! \Delta'_i \hat{n}_i ) \nonumber \\
        &+& \! \sum_i   \! \left[ V^\text{ex}_{i,i+1} ( \hat{\sigma}_i^+ \hat{\sigma}_{i+1}^- \! + \! \text{H.c.} ) \! + \! V_{i,i+1} \hat{n}_i \hat{n}_{i+1} \right], \quad
    \label{eq:HNN}
\end{eqnarray}
where $V_{i,i+1}=V_{i,i+1}^{\uparrow\uparrow}+V_{i,i+1}^{\downarrow\downarrow}$ accounts for the total nearest-neighbor vdW interaction strength.

From an experimental perspective, the regime considered in our work can be realized by tuning the inter-atomic spacing and principal quantum numbers. Since the vdW interaction scales as $V_{i,i+1} \propto n^{11}/a^6$ and the dipolar exchange scales as $V^\text{ex}_{i,i+1} \propto n^4/a^3$, the ratio $V_{i,i+1}/V^\text{ex}_{i,i+1}$ can be significantly enhanced by decreasing the lattice spacing $a$ within the validity range of atomic perturbation theory (i.e., beyond the wavefunction overlap limit and avoiding F\"orster resonances) and choosing higher principal quantum numbers $n$ \cite{reinhardLevelShiftsRubidium2007,walkerConsequencesZeemanDegeneracy2008,PhysRevLett.120.113602}. This makes it possible to reach a regime where the nearest-neighbor interaction dominates over all other energy scales relevant to the dynamics ($V_{i,i+1} \gg |\Delta'|, |\Omega|, V^\text{ex}_{i,i+1}$), while $\Delta'_i$ remains finite by adjusting $\Delta$, preventing trivial polarization. In this strong-interaction limit, the system is effectively approximated by $V_{i,i+1} \to \infty$, enforcing the Rydberg blockade where simultaneous excitations on adjacent sites (e.g., $\ket{\dots \uparrow \uparrow \dots}$) are strictly excluded from the low-energy subspace, which is formally defined by the global projector $\mathcal{P}=\prod_{i}(1-\hat{n}_i \hat{n}_{i+1})$. Assuming a uniform lattice with translational invariance ($V^{\rm ex}_{i,i+1} \equiv V^{\rm ex}$, $\Delta'_i \equiv \Delta'$), we define the dimensionless parameters $\lambda = \Omega/(2V^{\rm ex})$ and $h = \Delta'/(2V^{\rm ex})$. By setting the energy scale to $2V^{\rm ex}$, we obtain the final effective Hamiltonian:
\begin{eqnarray}
%    \begin{split}
        \hat{H}_{\text{eff}} & =&\sum_i \bigg[ \frac{1}{2} P_{i-1} (\hat{\sigma}_i^+ \hat{\sigma}_{i+1}^- + \text{H.c.}) P_{i+2} \nonumber \\ & &\quad+ \lambda P_{i-1} \hat{\sigma}_i^x P_{i+1} + h P_{i-1} \hat{n}_i P_{i+1} \bigg],
   % \end{split}
    \label{eq:Heff}
\end{eqnarray}
where the local projectors $P_i = 1 - \hat{n}_i$ enforce the blockade constraint by restricting dynamics to the subspace where neighboring sites are in the $\ket{\downarrow}$ state. To maintain structural consistency with the non-commuting Rabi and exchange terms while manifesting the locality of the constraint, we represent the detuning term as $h \sum_i P_{i-1} \hat{n}_i P_{i+1}$. In the strong-blockade regime, the effective detuning $h$ incorporates interaction-induced background shifts distinct from the bare detuning $\Delta$, thereby tuning relative energies strictly within this constrained subspace. Complementing the exchange-dominated regime of Ref. \cite{schollMicrowaveEngineeringProgrammable2022}, our setup accesses the strong-blockade limit by tuning and, thereby expanding the accessible parameter space. This ensures the dynamics remain confined to the constrained subspace. A detailed benchmark against the full microscopic model, along with the specific interaction parameters calculated via the ARC package \cite{SIBALIC2017319}, is presented in Appendix~\ref{app:validation}.

\begin{figure}[t]
    \centering
    \includegraphics[width=\columnwidth]{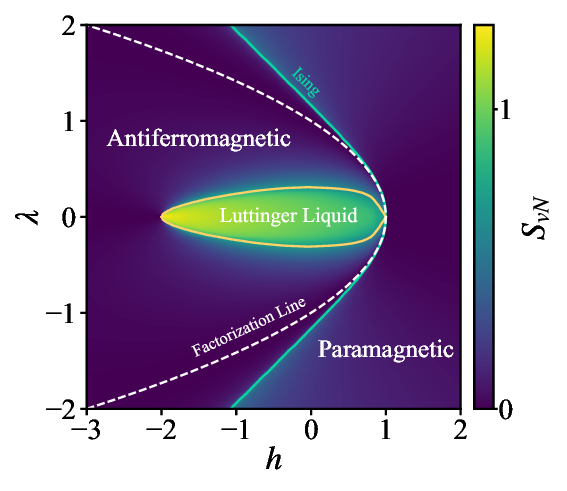}
    % 让图标题居中（\centering 对后续内容生效，故标题会随图片一同居中）
    \caption{\label{fig:Fig2} Phase diagram showing the ground state entanglement entropy $S_{vN}$ of $H_{\text{eff}}$ in the $(h,\lambda)$ parameter plane for bond dimension $M=20$. The distinct entanglement signatures delineate three quantum phases: the yellow-green critical LL phase, the dark-shaded blue AFM phase, and the slightly lighter PM phase. The cyan solid curve marks the Ising transition, the yellow line separates the AFM and LL phases, and the white dashed line indicates the factorization line.}
\end{figure}

The effective model (\ref{eq:Heff}) describes a one-dimensional constrained quantum spin chain with competing transverse $\lambda$ and longitudinal $h$ fields, supplemented by an explicit XY spin-exchange interaction. Rather than a simple spin chain, Eq.~\eqref{eq:Heff} provides a unified framework that interpolates between two well-established limiting cases: first, in the absence of exchange interactions ($V^\text{ex}=0$), it reduces to the PXP model with detuning~\cite{fendleyCompetingDensitywaveOrders2004,PhysRevB.99.161101,bluvsteinControllingQuantumManybody2021} and exhibits confinement-deconfinement transitions~\cite{zhangQuantumManybodyScars2023,chengTunableConfinementdeconfinementTransition2022,zhangYangleeEdgeSingularity2025}; second, in the absence of local Rabi driving ($\Omega=0$), the model reduces to the folded XXZ chain~\cite{alcarazExactlySolvableConstrained1999,zadnikFoldedSpin122021,zadnikFoldedSpin122021a,pozsgayIntegrableSpinChain2021}.
In this limit, the Hamiltonian exhibits exact $U(1)$ conservation of particle number. Together with an emergent $U(1)$ conservation of the domain-wall number~\cite{bastianelloFragmentationEmergentIntegrable2022}, imposed by the NN blockade constraint, the model displays Hilbert-space fragmentation and emergent integrability, supporting an LL phase~\cite{verresenStableLuttingerLiquids2019}.

The derivation of Eq.~\eqref{eq:Heff} relies on a direct first-order projection of the microscopic dynamics onto the low-energy subspace defined by the strong vdW blockade, where the single-body Rabi drive and the two-body dipolar exchange compete on equal footing.
For completeness, we note that the mathematical structure of this effective Hamiltonian also encompasses the effective description of the tilted Ising chain in the strong-coupling limit, as detailed in Appendix~\ref{app:derivation}, establishing a generalized connection between these paradigmatic lattice models. 

We perform numerical simulations using a suite of many-body methods tailored to different physical regimes. Exact diagonalization (ED)~\cite{sandvikComputationalStudiesQuantum2010b,10.21468/SciPostPhys.2.1.003}  is performed for system sizes up to
$L = 30$. Ground-state properties of larger finite systems are obtained using the density matrix renormalization group (DMRG)~\cite{whiteDensityMatrixFormulation1992,whiteDensitymatrixAlgorithmsQuantum1993,schollwockDensitymatrixRenormalizationGroup2005,schollwockDensitymatrixRenormalizationGroup2011,fishmanITensorSoftwareLibrary2022} with open boundary conditions (OBCs), a maximum bond dimension of 500, and up to 30 sweeps, with a truncation error below $10^{-12}$ and an energy convergence of $10^{-8}$. Access to the thermodynamic limit is achieved using the variational uniform matrix product state (VUMPS) method~\cite{zauner-stauberVariationalOptimizationAlgorithms2018}, which uses a 2-site unit cell with a bond dimension up to $M=256$, with a truncation error below $10^{-12}$ while reaching an energy convergence criterion of $10^{-7}$.

Fig.~\ref{fig:Fig2} presents the ground-state phase diagram. The color scheme represents $S_{vN} = -\mathrm{Tr}(\rho_r \ln \rho_r)$, where $\rho_r$ denotes the reduced density matrix for half of the chain. Three distinct quantum phases are clearly resolved: a yellow-green region representing finite $S_{vN}$ at weak driving corresponds to the critical LL phase; the surrounding dark-shaded blue AFM region and the slightly lighter dark-shaded blue PM region both exhibit low, area-law entanglement. The cyan solid curve marks the continuous Ising transition between the AFM and PM phases, while the yellow solid line delineates the boundary between the AFM and LL phases, where the sharp rise in $S_{vN}$ signals the quantum melting of the AFM order into the critical LL phase. The white dashed line passing through the AFM region denotes the exact factorization line. Section~\ref{sec:results} offers an in-depth analysis of these phases and transitions.

%\section{RESULTS\label{RESULTS}}

\section{RESULTS}
\label{sec:results}
\begin{figure}[t]
    \includegraphics[width=\columnwidth]{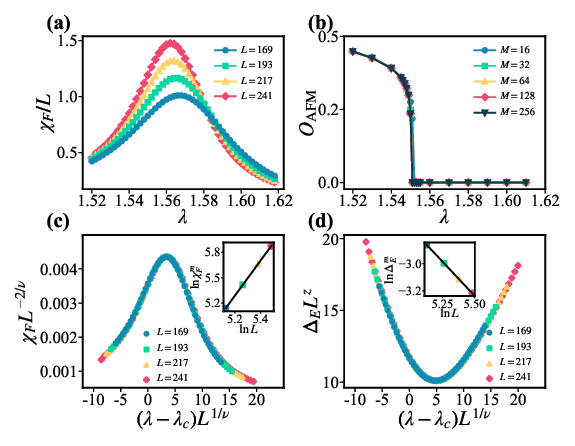}
    \caption{
        \label{fig:ising_transition}
        Phase transition and finite-size scaling at $h = -0.5$. Results are obtained using finite-size DMRG for various system sizes $L$ and VUMPS for the infinite system with varying bond dimensions $M$.
        (a) Fidelity susceptibility per site $\chi_F/L$ as a function of the Rabi drive $\lambda$.
        (b) $O_{\text{AFM}}$ as a function of $\lambda$ demonstrates a continuous vanishing behavior across the transition.
        (c) Finite-size scaling collapse of the fidelity susceptibility, utilizing the critical parameters $\lambda_c = 1.55$ and $\nu = 0.97 \pm 0.08$.
        The inset shows the power-law scaling of the peak value $\chi_F^{m}$ with system size $L$ on a log-log scale.
        (d) Corresponding data collapse for the neutral excitation gap, yielding a dynamical exponent $z = 1.01 \pm 0.03$ (with $\nu=1.00$ fixed).
        The inset depicts the gap scaling at the pseudocritical points $\lambda_{m}(L)$.
    }
    \makeatletter
    \def\@currentlabel{\thefigure(a)}\label{fig:ising_transition_a}
    \def\@currentlabel{\thefigure(b)}\label{fig:ising_transition_b}
    \def\@currentlabel{\thefigure(c)}\label{fig:ising_transition_c}
    \def\@currentlabel{\thefigure(d)}\label{fig:ising_transition_d}
    \makeatother
\end{figure}
%\subsection{Ising Transition\label{ising_transition}}

\noindent{\color{blue} \it{Ising Transition}.}- We first investigate the phase transition resulting from the competition between the interaction-stabilized AFM order and the quantum fluctuations associated with the Rabi drive. To independently locate the phase boundary without relying solely on the behavior of the order parameter, we compute the fidelity susceptibility~\cite{youFidelityDynamicStructure2007,chenIntrinsicRelationGroundstate2008,yuFidelitySusceptibilityDiagnostic2022}:
\begin{equation}
    \chi_F(\lambda) = -2\lim_{\delta\lambda \to 0} \frac{\ln|\langle \psi_\text{g}(\lambda) | \psi_\text{g}(\lambda + \delta\lambda) \rangle|}{(\delta\lambda)^2},
    \label{eq:chiF}
\end{equation}
which quantifies the sensitivity of the ground state $|\psi_\text{g}(\lambda)\rangle$ to infinitesimal changes in the control parameter $\lambda$. Numerical evaluation of this quantity for systems of varying linear sizes $L$ reveals that $\chi_F/L$ exhibits a pronounced maximum (Fig.~\ref{fig:ising_transition_a}). We denote the location of the maximum for each finite size $L$ as the pseudocritical point $\lambda_{m}(L)$, anticipating that the sequence $\{\lambda_m(L)\}$ will converge to a well-defined critical value $\lambda_c$ as $L\rightarrow \infty$.

Complementary to this information-theoretic metric, we compute the AFM order parameter $O_{\text{AFM}} = L^{-1} \sum_{j} (-1)^j \langle \hat{\sigma}^z_j \rangle$ using VUMPS directly in the thermodynamic limit. As shown in Fig.~\ref{fig:ising_transition_b}, $O_{\text{AFM}}$ is clearly nonzero for small Rabi drive strength $\lambda$, signaling robust antiferromagnetic order. Upon increasing $\lambda$ increases, the order parameter decreases smoothly and vanishes continuously at $\lambda_c=1.55$.

We now turn to a finite-size scaling analysis of the fidelity susceptibility and excitation gap to extract critical exponents, which will help substantiate the continuous nature of the transition. The peak value $\chi_F^{m}$ of the fidelity susceptibility grows with system size as a power law. Fitting $\chi_F^{m} \propto L^{2/\nu}$~\cite{PhysRevLett.103.170501} (see inset of Fig.~\ref{fig:ising_transition_c}) yields the correlation length exponent $\nu = 0.97 \pm 0.08$. Using this value of $\nu$ together with the estimated critical point $\lambda_c$, we obtain a striking data collapse of $\chi_F(\lambda, L)$ for different system sizes onto a single universal scaling function $\mathcal{F}_\chi$, as shown in Fig.~\ref{fig:ising_transition_c}. This collapse is fully consistent with the finite-size scaling ansatz~\cite{guFidelitySusceptibilityScaling2008,albuquerqueQuantumCriticalScaling2010}:
\begin{equation}
    \chi_F(\lambda, L) = L^{2/\nu} \mathcal{F}_{\chi}\big[ L^{1/\nu} (\lambda - \lambda_c) \big].
    \label{eq:chiF_scaling}
\end{equation}

To further test the consistency of the extracted critical parameters, we analyze the closing of the neutral excitation gap, $\Delta_E = E_1 - E_0$.
At the critical point, scaling theory predicts that this gap should close algebraically with system size as $\Delta_E \sim L^{-z}$, where $z$ is the dynamical critical exponent.
By fixing $\lambda_c$ and $\nu$ from the fidelity analysis, the gap data collapse onto a universal curve [Fig.~\ref{fig:ising_transition_d}], yielding a dynamical critical exponent $z = 1.01 \pm 0.03$. These results indicate that the phase transition between the AFM and PM phases belongs to the (1+1)D Ising universality class, with critical exponents $\nu \approx 1 $ and $z \approx 1$.

\noindent{\color{blue} \it{ Luttinger liquid phase}.}- We now turn to the critical regime emerging at weak driving.
Starting from the AFM phase at intermediate $\lambda$, reducing the Rabi drive induces a continuous melting of crystalline order.
The resulting critical state is identified as a floating LL phase~\cite{raderFloatingPhasesOneDimensional2019,zhangProbingQuantumFloating2025,soto-garciaNumericalInvestigationQuantum2025}.
This phase corresponds to the quantum melting of the commensurate $\mathbb{Z}_2$ solid.
As $\lambda$ decreases, quantum fluctuations overcome the pinning potential imposed by the blockade constraint, leading to the proliferation of domain walls.
In this regime, discrete translational symmetry is restored, and density-density correlations exhibit algebraic decay with incommensurate wave vectors that vary continuously with system parameters (See Appendix~\ref{app:correlations} for details).

\begin{figure}[t]
    \centering
    \includegraphics[width=\columnwidth]{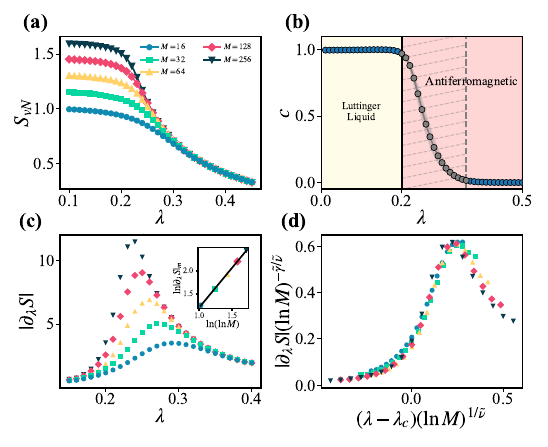}
    \caption{\label{fig:Fig4}
        Entanglement properties at fixed detuning $h = -0.5$ for various bond dimensions $M$. (a) $S_{vN}$ as a function of driving strength $\lambda$. (b) Effective central charge $c$ versus $\lambda$ extracted via finite-entanglement scaling. (c) Magnitude of the derivative $|\partial_\lambda S_{vN}|\equiv|dS_{vN}/d\lambda|$ versus $\lambda$. The inset shows the linear scaling of the peak maxima $\ln |\partial_\lambda S_{vN}|_m$ and their positions versus $\ln(\ln M)$. (d) Scaling collapse of $|\partial_\lambda S_{vN}|$ following the ansatz of Eq.~\eqref{eq:ansatz}, achieved with critical parameters $\lambda_c = 0.20$, $\tilde{\nu} = 1.00$, and $\tilde{\gamma} = 1.74$.
    }
    \makeatletter
    \def\@currentlabel{\thefigure(a)}\label{fig:LL_critical_a}
    \def\@currentlabel{\thefigure(b)}\label{fig:LL_critical_b}
    \def\@currentlabel{\thefigure(c)}\label{fig:LL_critical_c}
    \def\@currentlabel{\thefigure(d)}\label{fig:LL_critical_d}
    \makeatother
\end{figure}

To unambiguously identify this phase and distinguish it from the AFM order, we analyze the scaling behavior of the entanglement entropy $S_{vN}$. The scaling relation is given by \cite{tagliacozzoScalingEntanglementSupport2008,pollmannTheoryFiniteentanglementScaling2009}
\begin{equation}
    S_{vN} \sim \frac{c}{6} \ln \xi(M),
    \label{eq:Svn_scaling}
\end{equation}
where $c$ is the central charge of the underlying conformal field theory, and $\xi(M)$ denotes the effective correlation length. In infinite MPS simulations, while the physical correlation length diverges in the LL phase, the finite bond dimension $M$ introduces a cutoff length $\xi(M)$~\cite{PhysRevB.55.2164}, extracted from the eigenvalues of the transfer matrix.

As shown in Fig.~\ref{fig:LL_critical_a}, the entropy displays a sharp contrast between the two regimes: for $\lambda \gtrsim 0.3$, $S_{vN}$ saturates to a constant value independently of the bond dimension $M$, satisfying area law for a ground state; conversely, for $\lambda \lesssim 0.2$, it exhibits a persistent logarithmic divergence with increasing $M$, a hallmark of critical systems.

To delineate the extent of the LL phase, we extract the effective central charge $c(\lambda)$ using finite-entanglement scaling.
Fig.~\ref{fig:LL_critical_b} shows that $c \approx 1$ (within $5\%$ tolerance) over a finite interval of $\lambda$, identifying the LL regime.

The phase boundary is further characterized by the entanglement response
$\bigl| dS_{vN}/d\lambda \bigr|$.
As shown in Fig.~\ref{fig:LL_critical_c}, the response develops a pronounced peak whose height increases and whose width narrows with increasing bond dimension $M$, indicating a divergent susceptibility associated with the continuous
transition.
Importantly, the inset of Fig.~\ref{fig:LL_critical_c} reveals a clear power-law growth
of the peak maximum as a function of $\ln M$.

This behavior is naturally understood within the framework of finite-entanglement scaling. Since the entanglement entropy $S_{vN}$ scales logarithmically with the induced correlation length $\xi(M)$, the corresponding
entanglement response is governed by $\ln M$ rather than $M$ itself.
This motivates the following finite-entanglement scaling ansatz:
\begin{equation}
    \left| \frac{dS_{vN}}{d\lambda} \right| = (\ln M)^{\tilde{\gamma}/\tilde{\nu}}\, \mathcal{S}\!\left[ (\ln M)^{1/\tilde{\nu}} (\lambda - \lambda_c) \right], \label{eq:ansatz}
\end{equation}
where $\mathcal{S}$ is a universal scaling function.
As demonstrated in Fig.~\ref{fig:LL_critical_d}, data collapse is obtained
with $\lambda_c = 0.20$, $\tilde{\nu} \approx 1.00$, and $\tilde{\gamma} \approx 1.74$,
providing evidence for a continuous quantum melting transition into the
LL phase.

\begin{figure}[t]
    \centering
    \includegraphics[width=\columnwidth]{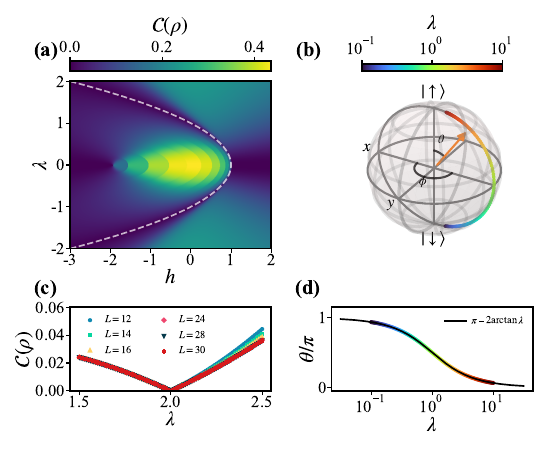}
    \caption{\label{fig:concurrence}
        (a) Nearest-neighbor concurrence $\mathcal{C}(\rho)$ in the $(h, \lambda)$ plane, obtained by ED for $L=24$.
        (b) Bloch sphere representation of the single-site state on the active sublattice; points are colored by $\lambda$.
        (c) Concurrence $\mathcal{C}(\rho)$ versus system size $L$ at three representative points along the factorization line at $h=-3.0$ (white dashed curve in (a)).
        (d) Polar angle $\theta$ of the Bloch vector as a function of $\lambda$; the solid black curve is the empirical fit $\theta(\lambda) = \pi - 2\arctan(\lambda)$.}

    \makeatletter
    \def\@currentlabel{\thefigure(a)}\label{fig:concurrence_a}
    \def\@currentlabel{\thefigure(b)}\label{fig:concurrence_b}
    \def\@currentlabel{\thefigure(c)}\label{fig:concurrence_c}
    \def\@currentlabel{\thefigure(d)}\label{fig:concurrence_d}
    \makeatother
\end{figure}

\noindent{\color{blue} \it{Exact Ground-State Factorization within the AFM Phase}.}-
Within the ordered AFM phase, we identify a one-dimensional trajectory along which quantum entanglement strictly vanishes, realizing a line of exact ground-state factorization (GSF)~\cite{roscildeEntanglementFactorizedGround2005,giampaoloTheoryGroundState2008, giampaoloSeparabilityGroundstateFactorization2009,yiCriticalityFactorizationHeisenberg2019}. Along this line, despite the strong competition between local driving and nonlocal exchange interactions, the ground state factorizes into an exact product state. This phenomenon is not a trivial consequence of suppressed quantum fluctuations due to large detuning, but instead arises from a precise quantum-interference mechanism in which the projected Rabi drive and dipolar exchange processes mutually cancel their entangling effects. The resulting factorized ground state thus provides a controlled realization of exact factorization in a constrained, interacting quantum system, closely related to factorized product states discussed in one-dimensional Rydberg crystals~\cite{zeybekFormationRydbergCrystals2025}.

To quantify the entanglement structure, we compute the nearest-neighbor concurrence $\mathcal{C}(\rho_{i,i+1})$, a standard measure of bipartite entanglement defined via the Hill-Wootters formula~\cite{hillEntanglementPairQuantum1997,woottersEntanglementFormationArbitrary1998}:
\begin{equation}
    \mathcal{C}(\rho) = \max\left\{0,\, r_{1} - r_{2} - r_{3} - r_{4}\right\}, \label{eq:concurrence}
\end{equation}
where $r_i$ are the square roots of the eigenvalues (in descending order) of the spin-flipped density matrix $R_{i,i+1} = \rho_{i,i+1} (\sigma^y_i \otimes \sigma^y_{i+1}) \rho_{i,i+1}^* (\sigma^y_i \otimes \sigma^y_{i+1})$.
A vanishing concurrence ($\mathcal{C} = 0$)  serves as a sufficient criterion for two-site separability.

We derive the exact analytical condition for factorization in Appendix~\ref{app:factorization_derivation}, obtaining the relation:
\begin{equation}
    h = 1 - \lambda^2.
    \label{eq:PE_line}
\end{equation}
Using exact diagonalization for system sizes up to $L=24$,
we map the concurrence across the $(h, \lambda)$ parameter plane.
As shown in Fig.~\ref{fig:concurrence_a}, the concurrence $\mathcal{C}$ vanishes identically along a smooth trajectory that coincides precisely with the analytical factorization line~\eqref{eq:PE_line}. Note that the surrounding stripe-like patterns originate from the Devil's staircase effect investigated in Ref.~\cite{PhysRevA.94.051603}. To exclude finite-size effects, we analyze the system size dependence at a representative point. Fixing $h = -3.0$, where factorization occurs at $\lambda = 2.0$, we plot $\mathcal{C}$ versus $\lambda$ for various system sizes in Fig.~\ref{fig:concurrence_c}. The concurrence approaches zero at the same value of $\lambda = 2.0$ for all system sizes, indicating that the separability persists in the thermodynamic limit.

Microscopically, the factorized ground state exhibits spontaneous breaking of translational symmetry into a dimerized pattern. The ground state takes the staggered product form
\begin{align}
    \ket{\psi_{\text{g}}} = & \bigotimes_k \ket{\downarrow}_{2k} \otimes \ket{\theta}_{2k+1},
\end{align}
where the even sublattice is frozen in the vacuum state $\ket{\downarrow}$ to satisfy the blockade constraint, while the active odd sublattice occupies a coherent single-site superposition
$\ket{\theta} = \cos\frac{\theta}{2}\ket{\uparrow} + e^{i\phi}\sin\frac{\theta}{2}\ket{\downarrow}$.
This local structure is illustrated on the Bloch sphere in Fig.~\ref{fig:concurrence_b}, where the state vectors of the active sites trace a smooth meridian as the driving strength is varied.
For $\lambda>0$,  the polar angle obeys the exact relation
\begin{equation}
    \theta(\lambda) = \pi - 2\arctan(\vert\lambda\vert),
    \label{eq:exct_lambda}
\end{equation}
with a fixed azimuthal phase $\phi=\pi$. For $\lambda<0$, $\phi$ shifts to $0$ by symmetry, while the factorization condition (\ref{eq:exct_lambda}) remains unchanged. As shown in Fig.~\ref{fig:concurrence_d}, the numerical results collapse onto this analytical curve, confirming that the factorized ground state is stabilized by an exact cancellation between local and nonlocal fluctuation channels.

\section{Conclusion}
In this work, we have mapped the complete ground-state phase diagram of a one-dimensional constrained Rydberg spin chain with competing local and nonlocal coherent processes. By considering finite detuning and coherent driving on equal footing, we extend this picture by providing a unified characterization of the zero-temperature phase diagram, which reveals an LL phase at weak driving (strong dipolar exchange), separated from the ordered phase by a continuous quantum melting transition.

Our analysis shows that the AFM order is destabilized through two distinct mechanisms controlled by the Rabi drive strength $\lambda$. At strong driving, the transition from the AFM phase to the polarized paramagnetic phase falls into the standard (1+1)D Ising universality class, consistent with the spontaneous breaking of discrete translational symmetry. At weak driving, by contrast, the commensurate AFM order melts continuously into an LL phase, realizing a quantum melting transition characterized by LL criticality.

In addition, we identify an exact ground-state factorization line embedded within the AFM phase. Along this line, the ground state is an exact product state, resulting from destructive interference between projected local driving and dipolar exchange processes. The existence of such a factorization line provides a concrete and analytically controlled example of vanishing entanglement in a strongly interacting constrained system. From an experimental perspective, this factorization line offers a natural entanglement-free reference point for calibrating coherent dipole-dipole couplings in programmable Rydberg atom arrays. Beyond its immediate utility, these findings uncover the potential of constrained Rydberg systems for high-fidelity state preparation in quantum information processing. Our work establishes this platform as a promising arena for uncovering novel quantum critical phenomena and exploring exotic phases of matter in future quantum simulation experiments.

\begin{acknowledgments}
    We acknowledge the National Science Foundation of China (Grant Nos. 12475033 and 12174194) for financial support.
\end{acknowledgments}

\section*{Data Availability}
The data that support the findings of this article are openly available~\cite{jiang_2026_18973130}. 
\appendix

\section{Numerical Benchmark of the Effective Hamiltonian (3)}
\label{app:validation}

To justify the validity of the effective Hamiltonian [Eq.~\eqref{eq:Heff}] and its constrained Hilbert space, we perform ED on a periodic chain of $L=12$ sites, enabling a direct comparison between its ground state and that of the full Rydberg Hamiltonian [Eq.~\eqref{eq:Hfull}].

Following the computational framework outlined in Ref.~\cite{SIBALIC2017319}, we determine the physical interaction coefficients for a Rydberg chain involving the states $\{|60S_{1/2}, m_j=1/2\rangle, |61P_{1/2}, m_j=-1/2\rangle\}$. The results are summarized in Table~\ref{tab:C3C6}. 

\begin{table}[ht]
\centering
\caption{van der Waals and dipole-dipole interaction coefficients for $^{87}$Rb Rydberg states.}
\label{tab:C3C6}
\begin{tabular}{l|c|c}
\hline
Coefficient & Description & Value \\
\hline
$C_3$ & Dipolar exchange (60S–61P) & $0.043\ \mathrm{GHz}\cdot\mu\mathrm{m}^3$ \\
$C_6^{\downarrow\downarrow}$ & vdW ($|60S, 60S\rangle$) & $138.88\ \mathrm{GHz}\cdot\mu\mathrm{m}^6$ \\
$C_6^{\uparrow\uparrow}$ & vdW ($|61P, 61P\rangle$) & $6.56\ \mathrm{GHz}\cdot\mu\mathrm{m}^6$ \\
\hline
\end{tabular}
\end{table}

By varying the lattice spacing $a$ over the experimentally accessible range $5.76$--$15\;\mu\text{m}$, we probe a broad range of interaction strengths. In the full model, the vdW interactions $V^{\uparrow\uparrow}_{i,i+l}$ and $V^{\downarrow\downarrow}_{i,i+l}$ scale as $(al)^{-6}$, while the resonant exchange coupling $V^{\text{ex}}_{i,i+l}$ scales as $(al)^{-3}$. Throughout the calculation, the single-body control parameters are fixed at $\Delta'/h_P = -0.05\,\text{MHz}$ and $\Omega/h_P = 0.05\,\text{MHz}$, here $h_P$ is the Planck constant.

\begin{figure}[t]
    \centering
    \includegraphics[width=\columnwidth]{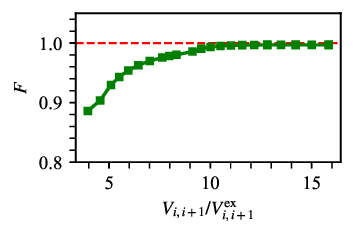}
    \caption{\label{fig:benchmark}    
     Ground-state fidelity $F = |\langle \psi_{\rm full} | \psi_{\rm eff} \rangle|$ as a function of $V_{i,i+1}/V^{\text{ex}}_{i,i+1}$, obtained via ED for a periodic chain of $L=12$. Parameters are set to $\Omega = 0.05$ and $\Delta' = -0.05$. The dashed line indicates $F=1$.}
\end{figure}

To quantify the accuracy of the effective description, we calculate the ground-state fidelity $F = |\langle \psi_{\rm full} | \psi_{\rm eff} \rangle|$ as a function of the dimensionless ratio $V_{i,i+1}/V^{\text{ex}}_{i,i+1}$ (where $V_{i,i+1} = V_{i,i+1}^{\uparrow\uparrow} + V_{i,i+1}^{\downarrow\downarrow}$). The benchmarking results are summarized in Fig.~\ref{fig:benchmark}. As illustrated, the fidelity increases monotonically with this ratio, demonstrating an asymptotic convergence toward unity as the system enters the strong-blockade regime ($V_{i,i+1} \gg |\Delta'|, |\Omega|, V^\text{ex}_{i,i+1}$). For a representative configuration with $a \approx 7\,\mu\text{m}$, the parameters are given by $V_{i,i+1}/h_P \approx 1.27\,\text{MHz}$  and $V^{\text{ex}}_{i,i+1}/h_P \approx 0.13\,\text{MHz}$, corresponding to $V_{i,i+1}/V^{\text{ex}}_{i,i+1} \approx 10$. At this typical experimental point, the fidelity exceeds $0.99$, confirming that the constrained effective Hamiltonian [Eq.~\eqref{eq:Heff}] provides an accurate description of the many-body ground state. The high overlap confirms that even in the presence of long-range interaction tails in the real system, the constrained effective Hamiltonian [Eq.~\eqref{eq:Heff}] provides a remarkably accurate description of the many-body ground state. This validates the use of the simplified model as a reliable foundation for the phase diagram analysis presented in this work.

\section{Connection to the Tilted Ising Model}
\label{app:derivation}

In this appendix, we discuss the relationship between the effective Rydberg Hamiltonian [Eq.~\eqref{eq:Heff}] and the microscopic tilted Ising model. While the effective model in the main text is derived from first-order dipolar exchange ($V^\text{ex}>0$), its mathematical structure is isomorphic to the effective description of a tilted Ising chain in the strong-coupling limit. Here, we derive this connection to highlight the generalized nature of the effective Hamiltonian.

Consider the tilted Ising Hamiltonian for a spin-1/2 chain governed by nearest-neighbor Ising interactions and local fields:
\begin{equation}
    \hat{H}_{\text{spin}} = J_{zz} \sum_j \hat{\sigma}_j^z \hat{\sigma}_{j+1}^z + h_x \sum_j \hat{\sigma}_j^x + h_z \sum_j \hat{\sigma}_j^z,
    \label{eq:H_micro}
\end{equation}
where $J_{zz}$ is the Ising coupling, and $h_x, h_z$ are the transverse and longitudinal fields. To elucidate the blockade physics, we map the spin-1/2 operators to a lattice gas (Rydberg) basis by identifying $\ket{\uparrow}$ with a particle presence ($\hat{n}_j=1$) and $\ket{\downarrow}$ with the vacuum ($\hat{n}_j=0$).Using the relation $\hat{\sigma}_j^z = 2\hat{n}_j - 1$, the Hamiltonian in the Rydberg basis is expressed as:
\begin{equation}
    \hat{H} = H_0 + \hat{V},
    \label{eq:H_boson}
\end{equation}
where $H_0 = 4J_{zz} \sum_j \hat{n}_j \hat{n}_{j+1}$ represents the interaction energy, and $\hat{V} = h_x \sum_j \hat{\sigma}_j^x + (2h_z - 4J_{zz})\sum_j \hat{n}_j$ incorporates the transverse and effective longitudinal fields. Constant energy shifts have been omitted. We proceed in the strong coupling limit, $4J_{zz} \gg |h_x|, |2h_z - 4J_{zz}|$, where the unperturbed Hamiltonian $\hat{H}_0$ enforces a constraint by splitting the Hilbert space into a low-energy subspace $\mathcal{P}$ (states with no adjacent particles, $\hat{n}_j \hat{n}_{j+1} = 0$) and a high-energy subspace $\mathcal{Q}$ (states with adjacent pairs costing $\Delta E \approx 4J_{zz}$).

We employ a Schrieffer-Wolff transformation to project the dynamics onto $\mathcal{P}$. The first-order effective Hamiltonian, $\hat{H}^{(1)} = \mathcal{P} \hat{V} \mathcal{P}$, projects the transverse field and detuning onto the constrained subspace. Since the operator $\hat{\sigma}_j^x$ can only flip a spin if nearest neighbors are empty to remain in $\mathcal{P}$, this yields the standard PXP model:
\begin{equation}
    \hat{H}^{(1)} = h_x \sum_j P_{j-1} \hat{\sigma}_j^x P_{j+1} + \mu \sum_j P_{j-1} \hat{n}_jP_{j+1},
\end{equation}
with the renormalized detuning $\mu=2h_z - 4J_{zz}$. The correlated hopping term arises from second-order virtual processes \cite{linQuasiparticleExplanationWeakthermalization2017,bastianelloFragmentationEmergentIntegrable2022}, $\hat{H}^{(2)} = \frac{1}{2} \mathcal{P} [\hat{S}, \hat{V}] \mathcal{P}$. Physically, this describes a process where the transverse field $h_x$ creates a virtual excitation in $\mathcal{Q}$ (a state with occupied nearest-neighbor sites $\dots \uparrow_j \uparrow_{j+1} \dots$) at an energy cost of $4J_{zz}$, which immediately de-excites to a neighboring site via a second action of $h_x$. The effective amplitude for this tunneling process is given by second-order perturbation theory as $J_{\text{eff}} \approx \frac{(h_x)(h_x)}{0 - 4J_{zz}} = -\frac{h_x^2}{4J_{zz}}$. Summing over all sites and ensuring the blockade constraint is satisfied at the boundaries of the hopping pair, we obtain the XY interaction term:
\begin{equation}
    \hat{H}_{\text{hop}} = -\frac{h_x^2}{4J_{zz}} \sum_j P_{j-1} \left( \hat{\sigma}_j^+ \hat{\sigma}_{j+1}^- + \text{h.c.} \right) P_{j+2}.
\end{equation}
This establishes a mapping to the effective Hamiltonian in the main text with the coefficients:
\begin{equation}
    J_{\text{Ising}} \equiv -\frac{h_x^2}{2J_{zz}}, \quad \Omega \equiv h_x, \quad \Delta' \equiv 2h_z - 4J_{zz}.
\end{equation}
Note that the perturbative derivation from the Ising model relies on a repulsive Ising interaction ($J_{zz} > 0$) to enforce the blockade constraint. This requirement strictly restricts the resulting effective exchange coupling to be ferromagnetic ($J_{\text{Ising}} \equiv -h_x^2/2J_{zz} < 0$). In contrast, the resonant dipolar exchange mechanism considered in the main text allows for an antiferromagnetic sign ($V^\text{ex} > 0$), thereby realizing a broader parameter regime that includes the phase diagram discussed in this work.

\section{Correlation Function and Structure Factor}
\label{app:correlations}

To characterize the nature of the phase transition and distinguish between the commensurate AFM phase and the incommensurate LL phase, we calculate the density-density correlation function. To probe the long-range crystalline order, we subtract the square of the mean density rather than the product of local densities:

\begin{equation}
    C(r) = \langle \hat{n}_i \hat{n}_{i+r} \rangle - \bar{n}^2,
    \label{eq:corr_def}
\end{equation}
where $\bar{n} = \frac{1}{L} \sum_i \langle \hat{n}_i \rangle$ is the average particle density. The static structure factor $S(k)$ is given by
\begin{equation}
    S(k) = \sum_{r} e^{-ikr} C(r).
\end{equation}

\begin{figure}[t]
    \centering
    \includegraphics[width=\columnwidth]{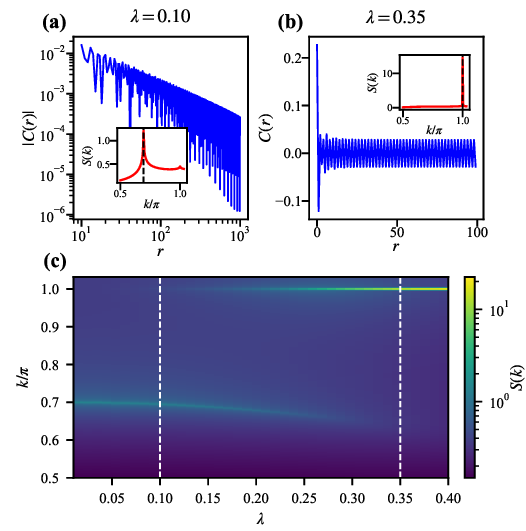}
    \caption{\label{fig:Fig7}
        Density-density correlations and structure factors from LL to AFM at $h=-0.5$.
        (a) Log-log plot of the correlation magnitude $|C(r)|$ in LL phase ($\lambda = 0.10$), demonstrating power-law decay characteristic of the LL phase. The inset shows the structure factor $S(k)$ peaking at an incommensurate wavevector.
        (b) Correlation $C(r)$ in the AFM phase ($\lambda = 0.35$), showing non-decaying oscillations indicative of long-range order. The inset displays a sharp Bragg peak at $k_{\text{C}}=\pi$.
        (c) Evolution of the structure factor $S(k)$ as a function of $\lambda$. The white dashed lines indicate the parameter values used in panels (a) and (b), where the AFM peak at $k_{\text{C}}=\pi$ fades away while an incommensurate mode emerges at $k_{\text{IC}} \approx 0.7\pi$.}
\end{figure}

The numerical results, obtained at a fixed transverse field $h=-0.5$, are summarized in Fig.~\ref{fig:Fig7}.
In the LL phase (e.g., $\lambda = 0.10$, Fig.~\ref{fig:Fig7}(a)), the system exhibits quasi-long-range order.
The magnitude of the correlations $|C(r)|$ follows an algebraic decay law, $C(r) \sim |r|^{-\eta}$, which appears as a straight line on a log-log scale.
The corresponding structure factor reveals a distinct peak at an incommensurate wavevector $k_{\text{IC}} \approx 0.7\pi$, confirming the floating nature of this phase.

In contrast, within the AFM regime (e.g., $\lambda = 0.35$, Fig.~\ref{fig:Fig7}(b)), the correlations $C(r)$ show persistent oscillations without decay, indicative of true long-range order.
This period-2 modulation leads to a sharp Bragg peak at $k_{\text{C}}=\pi$ in the structure factor (inset).

Fig.~\ref{fig:Fig7}(c) illustrates the evolution of the structure factor $S(k)$ as a function of the frustration parameter $\lambda$.
Instead of a continuous shift of the modulation wavevector, we observe spectral weight transfer in momentum space between competing orders.
Starting from the AFM phase (large $\lambda$), the commensurate peak is pinned at $k_{\text{C}}=\pi$.
As $\lambda$ decreases, the intensity of the AFM peak diminishes and eventually vanishes.
Simultaneously, a separate incommensurate peak emerges at a wavevector $k_{\text{IC}} \approx 0.7\pi$.
The coexistence and competition between the commensurate and incommensurate modes, characterized by a gradual transfer of spectral weight in $S(k)$, suggests a unique melting mechanism where the floating phase emerges without a continuous drift of the primary wavevector.

\section{Exact Derivation of the Factorization Ground State Line}
\label{app:factorization_derivation}

In this appendix, we present a complete and self-contained derivation of the exact condition under which the ground state of the effective Rydberg Hamiltonian becomes fully factorized, that is, a product state across all lattice sites. This remarkable simplification arises not from the absence of interactions, but rather from a precise cancellation between competing quantum processes: local Rabi driving and non-local exchange mediated by constrained hopping. The resulting destructive interference eliminates all entanglement, yielding a classical-like dimerized configuration that nonetheless emerges from a genuinely quantum many-body Hamiltonian.

We begin with the effective Hamiltonian obtained in the main text, Eq.~\eqref{eq:Heff}. Our goal is to identify a product state $\ket{\psi_\text{g}}$ that satisfies this constraint and is an exact eigenstate of $\hat{H}_{\text{eff}}$.

Given the alternating nature of the blockade, it is natural to consider a symmetry-broken ansatz that distinguishes even and odd sites. We therefore partition the chain into two sublattices: sublattice $A$, comprising even sites, and sublattice $B$, comprising odd sites. In the proposed ground state, all sites on sublattice $A$ are frozen in the vacuum state $\ket{\downarrow}$, while sites on sublattice $B$ carry coherent quantum superpositions. Explicitly, the trial state takes the form
\begin{equation}
    \ket{\psi_\text{g}} = \bigotimes_{k} \ket{\downarrow}_{2k} \otimes \ket{\psi_B}_{2k+1},
\end{equation}
with each active site described by a single-qubit state
\begin{equation}
    \ket{\psi_B} = \cos\frac{\theta}{2} \ket{\uparrow} + e^{i\phi} \sin\frac{\theta}{2} \ket{\downarrow}.
\end{equation}
This dimerized structure automatically respects the Rydberg blockade, as no two adjacent sites can both be excited.

However, for $\ket{\psi_\text{g}}$ to be an exact eigenstate, it is not sufficient to merely satisfy the static constraint; the dynamics generated by $\hat{H}_{\text{eff}}$ must also leave the state invariant up to a global phase. The key requirement is the stability of the supposedly inert vacuum sites on sublattice $A$. Although these sites are nominally empty, the Hamiltonian contains terms that could, in principle, generate excitations on them—provided the neighboring sites are in the vacuum. Crucially, such processes must cancel exactly for the product structure to survive.

To see how this cancellation occurs, consider a minimal three-site cluster consisting of a central site $A$ flanked by two active sites $B_L$ and $B_R$. An excitation on $A$ can be created through two distinct physical mechanisms, both conditioned on the neighbors being in $\ket{\downarrow}$. First, the projected Rabi term $\lambda P_{B_L} \hat{\sigma}_A^x P_{B_R}$ can flip the central spin from $\ket{\downarrow}$ to $\ket{\uparrow}$, with an amplitude proportional to $\lambda \braket{ \downarrow | \psi_B }_{B_L} \braket{ \downarrow | \psi_B }_{B_R}$. Second, the correlated hopping term—such as $\frac{1}{2} P_{B_L-1} \hat{\sigma}_{B_L}^- \hat{\sigma}_A^+ P_{B_R+1}$—can transfer an excitation from $B_L$ to $A$, provided $B_R$ remains in the vacuum; this contributes an amplitude proportional to $\braket{ \downarrow | \hat{\sigma}^- | \psi_B }_{B_L} \braket{ \downarrow | \psi_B }_{B_R}$. By symmetry, the contribution from the right neighbor is identical.

The total amplitude for creating an excitation at site $A$ is thus the sum of these two contributions. For the vacuum to remain stable, this total amplitude must vanish when projected onto $\ket{\uparrow}_A$. This leads to the interference condition
\begin{equation}
    \label{eq:interference}
    \lambda \left( \braket{ \downarrow | \psi_B } \right)^2 + \braket{ \downarrow | \hat{\sigma}^- | \psi_B } \braket{ \downarrow | \psi_B } = 0.
\end{equation}
Substituting the explicit expressions $\braket{ \downarrow | \psi_B } = e^{i\phi}\sin(\theta/2)$ and $\braket{ \downarrow | \hat{\sigma}^- | \psi_B } = \cos(\theta/2)$, the condition becomes
\begin{equation}
    \lambda e^{i\phi} \sin\frac{\theta}{2} + \cos\frac{\theta}{2} = 0.
\end{equation}
Solving for the mixing angle yields $\cot(\theta/2) = -\lambda e^{i\phi}$. Since the physical parameters $\lambda$ and $h$ are real, we may choose a gauge such that the relative phase is fixed. For $\lambda > 0$, setting $\phi = \pi$ renders the right-hand side real and positive, leading to the geometric relation
\begin{equation}
    \label{eq:theta_lambda}
    \theta = \pi - 2 \arctan \lambda.
\end{equation}

With the vacuum sublattice stabilized by this destructive interference, the remaining dynamics reduce to independent single-site problems at sublattice $B$. Because the neighboring $A$ sites are locked in $\ket{\downarrow}$, the projectors $P_{A}$ act as identities on the active sites. Consequently, the effective Hamiltonian governing each $B$ site simplifies to
\begin{equation}
    \hat{H}_{\text{eff}}^{(B)} = \lambda \hat{\sigma}^x + h \hat{n} \doteq
    \begin{pmatrix}
        h       & \lambda \\
        \lambda & 0
    \end{pmatrix},
\end{equation}
written in the basis $\{\ket{\uparrow}, \ket{\downarrow}\}$. The state $\ket{\psi_B} = (\cos\frac{\theta}{2}, -\sin\frac{\theta}{2})^T$ (with $\phi = \pi$) must be an eigenvector of this matrix. Imposing the eigenvalue equation
\begin{equation}
    \begin{pmatrix} h & \lambda \\ \lambda & 0 \end{pmatrix}
    \begin{pmatrix} \cos\frac{\theta}{2} \\ -\sin\frac{\theta}{2} \end{pmatrix}
    = E
    \begin{pmatrix} \cos\frac{\theta}{2} \\ -\sin\frac{\theta}{2} \end{pmatrix},
\end{equation}
the second row immediately gives the single-site energy density $E = -\lambda \cot(\theta/2)$. Using the interference condition $\cot(\theta/2) = \lambda$ from Eq.~\eqref{eq:theta_lambda}, we find $E = -\lambda^2$. Substituting this result into the first row yields
\begin{equation}
    h \cos\frac{\theta}{2} - \lambda \sin\frac{\theta}{2} = -\lambda^2 \cos\frac{\theta}{2},
\end{equation}
which, after rearrangement and division by $\cos(\theta/2)$ (assuming $\theta \neq \pi$), gives
\begin{equation}
    h = \lambda \tan\frac{\theta}{2} - \lambda^2.
\end{equation}
But from $\cot(\theta/2) = \lambda$, we have $\tan(\theta/2) = 1/\lambda$, and thus
\begin{equation}
    h = 1 - \lambda^2.
\end{equation}

This final relation defines the exact line in parameter space $(h, \lambda)$ along which the ground state is fully factorized and given by the product state $\ket{\psi_\text{g}}$.

\bibliography{apssamp}% Produces the bibliography via BibTeX.
\end{document}